\documentclass[12pt]{article}
\usepackage{graphicx}

\begin{document}
\title{Density dependent relativistic hadron theory for hadron matter with inclusion of pentaquark $\Theta^+$
\thanks{Supported by Chinese Academy of Sciences Knowledge Innovation Project (KJCX2-SW-No2),
 National Natural Science Foundation of China(10435080) }\\}
\author{\small Xi-Guo Lee$^{1,2}$, Yuan Gao$^{2,3}$    }
\date{}
\maketitle
\begin{center}
$^{1}${\small Institute of Modern Physics,Chinese Academy of
Sciences, P.O.Box 31 Lanzhou 730000,China}\\
$^{2}${\small Center of Theoretical Nuclear Physics, National
Laboratory of Heavy Ion Collisions,\\P.O.Box 31 Lanzhou730000,China} \\
$^{3}${\small Graduate School of The Chinese Academy Of Sciences,
Beijing, 100049, China}
\end{center}

\begin{abstract}
The density dependent relativistic hadron field theory(DDRH) is
extended to inclusions of pentaquark $\Theta^+$.We investigate
In-medium properties of $\Theta^+$ and nucleons.As well as the
neutron and proton,the effective $\Theta^+$ mass decreases as the
baryon density increases,and remains larger than nucleons'.The
effective mass of $\Theta^+$ is
calculated,$M^*_{\Theta^+}\simeq0.71M_{\Theta^+}$ at normal
nuclear density with the fraction of $\Theta^+$ 0.2. The effective
masses of all baryon increase with the increasing fraction of
$\Theta^+$.The binding energy is also studied.We find the binding
energy is much larger than it in nuclear matter without
$\Theta^+$,and then discuss the stability of the system with
conclusions of pentaquark $\Theta^+$. It is shown that the system
becomes bounder added small fraction of  $\Theta^+$ .finally we
discuss the condition when the system is the most stable and
calculate the minimum of binding energy and the fractions  of all
baryons as functions of baryon density.
\end{abstract}




\section{Introduction}

\hskip 0.3in The possibility of the existence of pentaquark
$\Theta^+$$(uudd\bar{s})$has been theoretically discussed for many
years.In 1997, Diakonov, Petrov, and Polyakov ,in the framework of
the chiral soliton model, predicted a exotic baryon $\Theta^+$
with mass $M\sim1.53Gev$,$S = +1$,spin 1/2 and isospin
0\cite{Diakonov}. However,the subject has been thrust to the fore-
front during the past two years by the experimental discovery of
an exotic
baryon\cite{Nakano1,Nakano2,Barmin,Stepanyan,Barth,Kubarovsky,Barth,Aleev,Aratayn,Airapetian}.It
has $K^+ n$ quantum numbers(B=+1, Q=+1, S=+1), and its minimal
quark content should be $uudd \bar s$. The remarkable features of
the $\Theta^+$ are its small mass (1540MeV) and very narrow width
($<$25MeV)\cite{Nakano2}.The discovery of the exotic baryon
$\Theta^{+}$ with positive strangeness opens new possibilities of
forming exotic $\Theta^{+}$ hypernuclei as same as in the case of
negative strangeness $\Lambda$, $\Sigma$, $\Xi$ hypernuclei. D.
Cabrera et al have suggested that $\Theta^{+}$ could be bound in
nuclei and calculated self energy of the $\Theta^{+}$ pentaquark
in nuclei\cite{Cabrera}.they shown that the in-medium
renormalization of the pion in the two meson cloud of the
$\Theta^{+}$ leads to a sizable attraction, enough to produce a
large number of bound and narrow $\Theta^{+}$ states in nuclei.

On other hand,the relativistic mean field approximation have been
a widely used and successful approach to describe properties of
nuclear matter and finite nuclei\cite{Dobaczewski}.the Density
Dependent Relativistic Hadron  (DDRH) theory was introduced
previously as an effective field theory for isospin
nuclei\cite{Lenske,Fuchs}.In DDRH theory the meson-baryon
interactions is described by meson-nucleon vertices which are
functionals of the fermion field operators.The coupling
coefficients are not constants but field operators.DDRH theory
reduces to a Hartree description with density dependent coupling
constants similar to the initial proposal of Brockmann and Toki in
mean-field approximation and its development has been achieved a
significant progress in recent years\cite{Brockmann}.For including
the scalar isovector $\delta$ meson,DDRH is successfully applied
to asymmetric matter at extreme neutron-to-proton
ratios\cite{Hofmann}.

In the present paper we attempt to extend  DDRH theory to
inclusions of pentaquark $\Theta^+$ and investigate properties of
 matter with inclusions of $\Theta^+$.We study medium modifications within
the mean field theory and get the mass modifications of baryons
In-medium.We calculate the binding energy at different baryon
fractions.We find the system becomes more bound with inclusions of
$\Theta^+$ but when too much $\Theta^+$ added it becomes less
bound. The critical fraction of $\Theta^+$ is 0.56 in isospin
symmetric matter and it becomes lower in isospin symmetric matter
. The isospin effects on the effective mass and the binding energy
are also discussed .Finally,we calculate the minimum of binding
energy and the fractions  of all baryons as functions of baryon
density when  the system is the most stable.

\section{ The density dependent relativistic hadron field}
\hskip 0.3in The density dependent relativistic hadron field
theory has been proven to give a good description of nuclear
matter in bulk and of the properties of finite nucle\cite{Hofmann}
.Assuming the interaction between nucleons and mesons is similar
to $\Theta^+$ and meson, We start from the effective lagrangian
density which includes pentaquark $\Theta^{+}$
 \begin{equation}
\mathcal{L}=\mathcal{L}_{B}+\mathcal{L}_{M}+\mathcal{L}_{int},
\end{equation}
 \begin{equation}
\mathcal{L}_{B}=\sum_{i=N,\Theta^+}{\bar{\Psi}_i(i\gamma_{\mu}\partial^{\mu}+M_i)\Psi_i},
\end{equation}

\begin{eqnarray}
\mathcal{L}_{M}=\frac{1}{2}
\sum_{i=\sigma,\delta}{(\partial_{\mu}\Phi_i\partial^{\mu}\Phi_i-m_i
^2\Phi_i ^2)}-\frac{1}{2}\sum_{j=\omega,\rho}{(F^{(j)}_{\mu\nu}F^{
(j)^{\mu\nu}}-m_j^2A^{(j)}_{\mu}A^{(j){\mu}})}
\end{eqnarray}
\begin{eqnarray}
\mathcal{L}_{int}=
\sum_{i=N,\Theta^+}{(\bar{\Psi}_i\Gamma^i_{\sigma}(\rho)\Psi_i
\Phi_{\sigma}-\bar{\Psi}_i\Gamma^i_{\omega}(\rho)\gamma_{\mu}\Psi_iA^{(\omega)\mu})}
+\bar{\Psi}_{N}\Gamma^N_{\delta}(\rho)\tau_{a}\Psi_N\Phi_{\delta}^a
\cr-\bar{\Psi}_{N}\Gamma^N_{\rho}(\rho)\gamma_{\mu}\tau_{a}\Psi_N
A^{(\rho)a\mu}
\end{eqnarray}
where

\begin{eqnarray}
 F^{(j)}_{\mu\nu} =\partial_{\mu}A_\nu^{(j)} - \partial_{\nu}A_\mu^{(j)}
.
\end{eqnarray}
Here,$\mathcal{L}_{B}$ and $\mathcal{L}_{M}$ are the free baryonic
and the free mesonic Lagrangians, respectively, and interactions
are defined by $\mathcal{L}_{int}$.The meson fields are denoted by
$\phi_\sigma$,$\phi_\delta$, $A^\omega$,  and $A^\rho$, and their
masses by $m_{\sigma}$, $m_{\delta}$,$m_{\omega }$ and $m_{\rho}$,
respectively. The isospin Pauli matrices are written as
$\tau^{a}$, $\tau^{3}$ being the third component of $\tau^{a}$.
For exotic baryon $\Theta^+$, as we assume isospin $I=0$,
$\tau^{a}=0$, there is no coupling to $\rho$ and $\delta$
mesons.The main difference to DDRH theory in Ref.\cite{Hofmann} is
that the lagrangian density includes the interactions of
pentaquark $\Theta^+$ and mesons.In contrast to standard QHD
models\cite{walecka,Serot1,Serot2},the meson-baryon vertices
$\hat\Gamma_\alpha$ ($\alpha=\sigma,\omega,\delta, \rho$) are not
constant numbers but the baryon field operators $\Psi$ dependent.

In the mean field approximation(MFA),the meson field operators can
be replaced by their expectation values,which are classical
fields\cite{walecka,Serot1,Serot2}
\begin{eqnarray}
\phi_i\rightarrow\langle \phi_i \rangle=\phi_{i0}\\
A^i\rightarrow\langle A^i\rangle=A^i_0
\end{eqnarray}
The field equations are reduced to
\begin{eqnarray}
(-\nabla^2+m_{\sigma}^2)\Phi_{\sigma0}=\Gamma_{\sigma}^N
(\rho)\rho_{S}^{N}+\Gamma_{\sigma}^{\Theta^+}(\rho)\rho_{S}^{\Theta^+}
,\\
(-\nabla^2+m_{\delta}^2)\Phi_{\delta0}=\Gamma_{\delta}^N
(\rho)\rho^3_S,\\
(-\nabla^2+m_{\omega}^2)A^{(\omega)}_0=\Gamma_{\omega}^N(\rho)\rho^N
+\Gamma_{\omega}^{\Theta^+}(\rho)\rho^{\Theta^+}\\
(-\nabla^2+m_{\rho}^2)A^{(\rho)}_0=\Gamma_{\rho}^N(\rho)\rho^{3}
\end{eqnarray}
where the densities are values of the following ground state
expectation
\begin{eqnarray}
\rho_{S}^{i}=\langle\bar{\Psi}_i\Psi_i\rangle\\
 \rho^{i}=\langle\Psi_i^+\Psi_i\rangle
\\ \rho_S^3=\langle\bar{\Psi}_N
\tau_3\Psi_N\rangle=\rho_{S}^{P}-\rho_{S}^{N}
\\ \rho^3=\langle\Psi_N^+\tau_3\Psi_N\rangle=\rho^{P}-\rho^{N}
\end{eqnarray}

The Dirac equation, separated in isospin, is the only remaining
operator field equation
\begin{eqnarray}
[i\gamma_{\mu}\partial^{\mu}-\Gamma_{\omega}^N\gamma^0A_0^{\omega}-
\Gamma_{\rho}^N\gamma^0\tau_3A_0^{\rho}-(M_N-\Gamma_{\sigma}^N\Phi_{\sigma}-\tau_3\Gamma_{\delta}^N
\Phi_{\sigma})]\Psi_N=0
\end{eqnarray}
\begin{eqnarray}
[i\gamma_{\mu}\partial^{\mu}-\Gamma_{\omega}^{\Theta^+}\gamma^0A_0^{\omega}-(M_{\Theta^+}-\Gamma_
{\sigma}^{\Theta^+}\Phi_{\sigma})]\Psi_{\Theta^+}=0
\end{eqnarray}
The in-medium mass of nucleons and pentaquark $\Theta^+$ is then
given by
\begin{eqnarray}
M_P^*=M_N-\Gamma_{\sigma}^N(\rho)-\Gamma_{\delta}^N(\rho)\Phi_{\delta}\\
M_N^*=M_N-\Gamma_{\sigma}^N(\rho)+\Gamma_{\delta}^N(\rho)\Phi_{\delta}
\end{eqnarray}
The field equations can be further simplified assuming
translational invariance and neglecting the electromagnetic
field.Solutions of the stationary Dirac equation
\begin{eqnarray}
\left[\gamma_{\mu}k_b^{*\mu}-m_b^*\right]u_b^*(k)=0
\end{eqnarray}\\
The usual plane wave Dirac spinors is given by\cite{Bjorken}
\begin{eqnarray}
 u^*_{b}(k)=\sqrt {\frac{E^{*}_{b}+m^{*}_{b}}{2m^{*}_{b}}}
 \left(\begin{array}{ccc}
  1\\
 \frac{\sigma k^{*}_{\sigma}}{E^{*}_{b}+m^{*}_{b}}\end{array}\right)\chi_{b}~,
\end{eqnarray}

where $\chi_b$ is a two-component Pauli spinor and the index $b$
distinguishes between neutrons ,protons and pentaquark
$\Theta^{+}$.Due to the inclusion of the $\delta$ meson,The
effective mass differs for neutrons and protons.The kinetic
4-momenta $k_b^{*\mu}$ and the energy $E_b^*$ of the particle are
related by the in-medium on-shell condition,so it is easily to get
${k_b^{*}}^2={m_b^{*}}^2$.Integrating over all states $k\leq
k_{F_b}$ inside the Fermi sphere and introducing
$E_{F_b}=\sqrt{k_{F_b}^2+{m_b^*}^2}$ the scalar and vector
densities in infinite nuclear matter lead to
\begin{eqnarray}
\rho_b & = & \frac{2}{(2\pi)^3}\int_{|k|<k_{F_b}}d^3 k =
\frac{k_{F_b}^3}{3\pi^2}
\label{eq:RhoB}\\
\rho_b^s & = & \frac{2}{(2\pi)^3}\int_{|k|<k_{F_b}}d^3 k
\frac{m_b^*}
{E_b^*} \nonumber \\
         & = & \frac{m_b^*}{2\pi^2}\left[k_{F_b} E_{F_b}
           + {m_b^*}^2\ln\frac{k_{F_b}+E_{F_b}}{m_b^*}\right].
\end{eqnarray}

The density-momentum tensor is defined

\begin{eqnarray}
T^{\mu\nu} =
\sum_{i}{\frac{\partial\mathcal{L}}{\partial(\partial^\mu\Phi_i)}\partial_\nu\Phi_i}-g^{\mu\nu}\mathcal{L}
\cr=
\sum_{i}\bar{\Psi}_i\gamma_\mu\partial_\nu\Psi_i-g^{\mu\nu}(\frac{1}{2}m_\omega^2A_0^{\omega2}+
\frac{1}{2}m_\rho^2A_0^{\rho2}-
\frac{1}{2}m_{_\sigma0}^2\phi_{_\sigma0}^2-\frac{1}{2}m_\delta^2\phi_{\delta0}^2)
\end{eqnarray}
\begin{eqnarray}
\varepsilon=\langle T^{00}
\rangle=\frac{2}{(2\pi)^3}\int_0^{K_{Fn}}
d^3k(k^2+M_n^{*2})^\frac{1}{2}\cr+\frac{2}{(2\pi)^3}\int_0^{K_{Fp}}
d^3k(k^2+M_p^{*2})^\frac{1}{2}\cr+\frac{2}{(2\pi)^3}\int_0^{K_{F\Theta^+}}
d^3k(k^2+M_\Theta^{+*2})^\frac{1}{2}\cr+\frac{1}{2} \left[
m_{\sigma}^2\Phi_{\sigma}^2+ m_{\delta}^2\Phi_{\delta}^2 +
m_{\omega}^2{A^{(\omega)}_0}^2+ m_{\rho}^2{A^{(\rho)}_0}^2\right]
\end{eqnarray}
Here the meson-baryon vertices $\hat\Gamma_\alpha$
($\alpha=\sigma,\omega,\delta, \rho$) are
 dependent on the baryon field operators $\Psi$ rather than constant numbers. Relativistic
covariance requires that the vertices are functions
$\hat\Gamma_{\alpha}(\hat\rho)$ of Lorenz-scalar bilinear forms
$\hat\rho(\bar\Psi, \Psi)$ of the field operators.In mean-field
approximation they reduce to density dependent coupling. Hofmann
got a rational approximation as follows by mapping DB
calculations\cite{Hofmann}

\begin{eqnarray}
\Gamma_{\alpha}^N(\rho) = a_{\alpha}\left[
                        \frac{1+b_{\alpha}(\frac{\rho}{\rho_0}
                        +d_{\alpha})^2}
                             {1+c_{\alpha}(\frac{\rho}{\rho_0}
                             +e_{\alpha})^2}\right]
\end{eqnarray}

\begin{table}[h]
\begin{center}
\caption{The DDRMF model parameter sets are taken from
Ref.\cite{Hofmann}} \vspace{0.1cm}
\begin{tabular}{|c|c|c|c|c|c|c|c |}
\hline\hline

meson $\alpha$   & $\sigma$ & $\omega$ & $\delta$ & $\rho$\\
\hline $m_{\alpha}$[MeV]& 550      & 783      & 983      & 770
\\ \hline
$a_{\alpha}$     & 13.1334 & 15.1640 & 19.1023 & 12.8373 \\
$b_{\alpha}$     &  0.4258 &  0.3474 &  1.3653 &  2.4822 \\
$c_{\alpha}$     &  0.6578 &  0.5152 &  2.3054 &  5.8681 \\
$d_{\alpha}$     &  0.7914 &  0.5989 &  0.0693 &  0.3671 \\
$e_{\alpha}$     &  0.7914 &  0.5989 &  0.5388 &  0.3598 \\ \hline
\multicolumn{5}{c}{$\rho_0 = 0.16$ [$fm^{-3}$]} \\
\end{tabular}
\end{center}
 \end{table}

The theta-meson coupling constants can be determined with the
quark meson model developed in Ref.\cite{Pierre}.For isospin
$I=0$,we only consider $\sigma$ and $\rho$ meson.The equation of
motion for meson field operators are as follows
\begin{eqnarray}
   \partial_{\mu}\partial ^{\mu}
   \hat{\sigma}+m_{\sigma}^{2}\hat{\sigma}=g_{\sigma}^{q}\bar{q}q,\\
   \partial_{\mu}\partial ^{\mu}
   \hat{\omega}^{\nu}+m_{\omega}^{2}\hat{\omega}^{\nu}=g_{\omega}^{q}\bar{q}\gamma^{\nu}q,
   \end{eqnarray}
where $g_{\sigma}^{q}$ and $g_{\omega}^{q}$ are the quark-meson
coupling constants  for $\sigma$ and $\omega$,respectively. The
mean fields are defined as the expectation values
\begin{eqnarray}
\langle A|\hat{\sigma}(t,r)|A \rangle=\sigma(r),\\
\langle A|\hat{\omega}^{\nu}(t,r)|A
\rangle=\delta(\nu,0)\omega(r),
\end{eqnarray}
Where $|A\rangle$ is  the ground state of the nucleus and (t,r)
are the coordinates in the rest frame of the nucleus. In the mean
field approximation the sources are the sum of the sources
   created by each nucleon - the latter interacting with the meson
   fields. Thus
   \begin{eqnarray}
 \bar{ q}q(t,r)= \sum_{i=1, A}\langle \bar{ q}q(t,r)
 \rangle_{i},
  \\
  \bar{q}\gamma^{\nu}q(t,r)= \sum_{i=1, A}\langle  \bar{q}\gamma^{\nu}q(t,r)
 \rangle_{i},
 \end{eqnarray}
 where $\langle...\rangle_{i}$ denotes the matrix element in the
 nucleon $i$ located at $\mathbf{\emph{R}}_{i}$ at time $t$.Finally for
the strange baryon,X.H. Zhong etc have got the
relations\cite{zhong}
\begin{eqnarray}
g_{\sigma}^{S}=\frac{n_{q}}{3}g_{\sigma}^{N}\Gamma_{S/B} ,
   g_{\omega}^{S}=\frac{n_{q}}{3}g_{\omega}^{N},
   g_{\rho}^{S}=g_{\rho}^{N}.
\end{eqnarray}
Where $n_{q}$ is the total number of valence $u$ and $d$ quarks in
the baryon S.\\ For $\Theta^+$,$n_{q}$=4.$\Gamma_{S/B}\approx1$
for all hyperons in practice\cite{Tsushima}.In our calculation We
use the relations as follows
\begin{eqnarray}
\Gamma_{\alpha}^{\Theta^+}(\rho)
=\frac{4}{3}\Gamma_{\alpha}^N(\rho)
\end{eqnarray}

 The baryon density is given by
\begin{eqnarray}
\rho=\rho^n+\rho^p+\rho^{\Theta^+}\cr=\frac{1}{(3\pi)^2}K_{Fn}^3
+\frac{1}{(3\pi)^2}K_{Fp}^3+\frac{1}{(3\pi)^2}K_{F\Theta^+}^3
\end{eqnarray}
Where $K_{Fn}$,$K_{Fp}$ and $K_{F\Theta^+}$ are the Fermi momenta
for n,p and $\Theta^+$,respectively. The baryon fraction is
\begin{eqnarray}
Y_i=\frac{\rho^i}{\rho^n+\rho^p+\rho^{\Theta^+}}, i=n,p,\Theta^+
\end{eqnarray}
The energy per baryon for multi-Theta matter is defined
\begin{eqnarray}
\frac{E}{B}=\frac{\varepsilon}{\rho}-Y_nM_n-Y_pM_p-Y_{\Theta^+}
M_{\Theta^+}
\end{eqnarray}

\section{Results and discussion}
\hskip 0.3in We first calculate the effective baryonic masses with
different $\Theta^+$ fractions in isospin symmetric matter and
isospin asymmetric matter.The results are presented in Fig.1(a)
and Fig.1(b).It is shown that the effective masses all decrease as
the baryon density increases,whether the nucleons or the
$\Theta^+$.The isospin effects on the effective  nucleon mass can
not be neglected due to the interaction between the scalar
isovector $\delta$,especially at high density.The
neutron-proton(n/p) effective mass splits.But the effects on the
$m_{\Theta^+}^*$ is negligible.The effective Theta masses always
decrease more slowly than the nucleons.The fraction $Y_{\Theta^+}$
can also affect on baryonic masses.At the same baryon density,the
baryonic masses all increase with more conclusion of pentaquark
$\Theta^+$.

Fig.2(a) shows the binding energy in isospin symmetry matter.We
plot the binding energy with different $\Theta^+$ fractions.Our
calculation show that the binding energy is sensitive to the
variation of the $\Theta^+$ fraction.The absolute minimum of
binding energy increases until the $\Theta^+$ fraction reaches
about 0.23,witch is up to nearly 24Mev at the baryon density
1.4$\rho_0$.But when the fraction of $\Theta^+$ reaches to
0.56,the system becomes not bound anymore.

The binding energy of the system without proton is presented in
Fig.2(b).The system with inclusion of a small fraction of
pentaquark $\Theta^+$ becomes stable.The absolute minimum reaches
the peak at the fraction 0.3.More $\Theta^+$ added,system becomes
less bound.When the percent of pentaquark $\Theta^+$ reaches to
half an whole,system becomes unstable again.It shows that the
binding is also sensitive to the isospin by comparing with the
curves in Fig.2(a).

Minimizing the $E/B$ with respect to $Y_{\Theta^+}$ and $Y_p$,we
got the minimum of the binding energy.The results are given in in
Fig.3.It is shown that The minimum of binding energy occurs at the
density about 1.5$\rho_0$.In low-density
region($0\leq\rho\leq1.5\rho_0$),the binding energy increases with
the increasing baryon density.The opposite dependence occurs in
hight-density region($\rho_0\geq1.5\rho_0$). Fig.3(b) displays the
fractions of neutron,proton and pentaquark $\Theta^+$ as
calculated by minimizing the energy.The curves display the
neutron-to-proton ratio is near 1:1.It is easily to comprehend
since the asymmetric nuclear matter is more stable than symmetric
matter.

In summary,we calculate the Multi-Theta matter could be bound in
the framework of density dependent relativistic hadron field
theory.It seems possible that there is a large fraction of
pentaquark $\Theta^+$.Whether nucleons or pentaquark
$\Theta^+$,it's effective mass in-medium is decreasing with the
increasing baryon density.But effective mass of pentaquark
$\Theta^+$ is larger than that of the ${N,\Theta^+}$.

\newpage

\begin{figure}
\begin{center}
{\includegraphics*[width=0.6\textwidth]{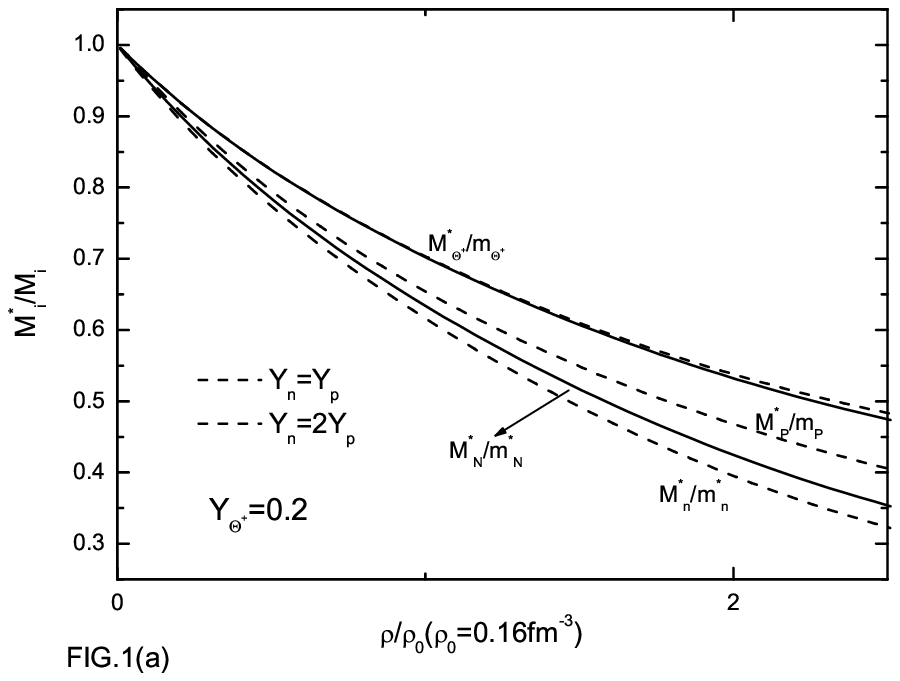}}
\end{center}
\begin{center}
{\includegraphics*[width=0.6\textwidth]{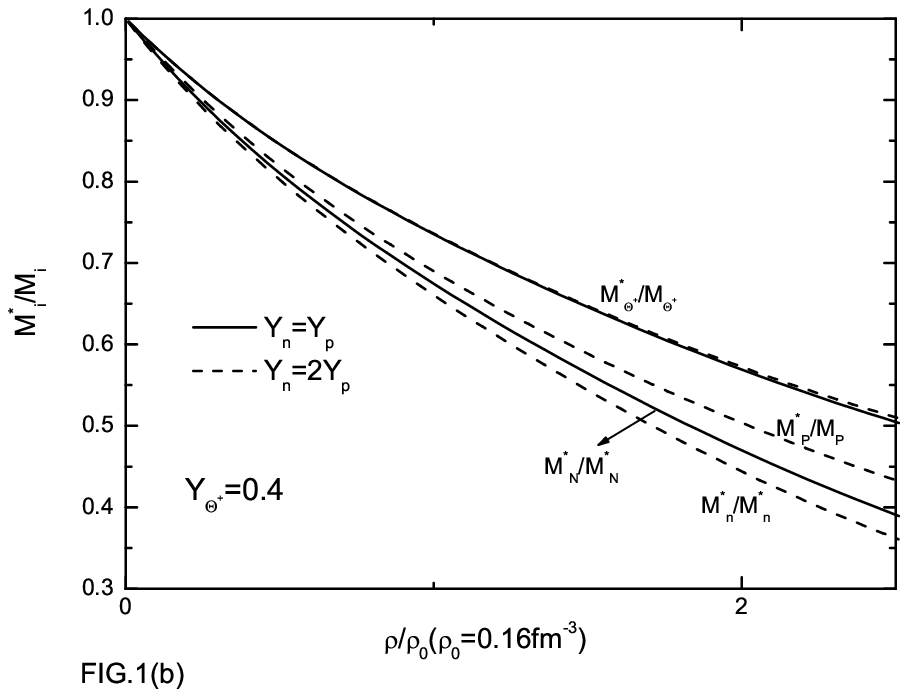}}
\end{center}
\caption{effective baryonic masses.}
\end{figure}
\begin{figure}

\begin{center}
{\includegraphics*[width=0.6\textwidth]{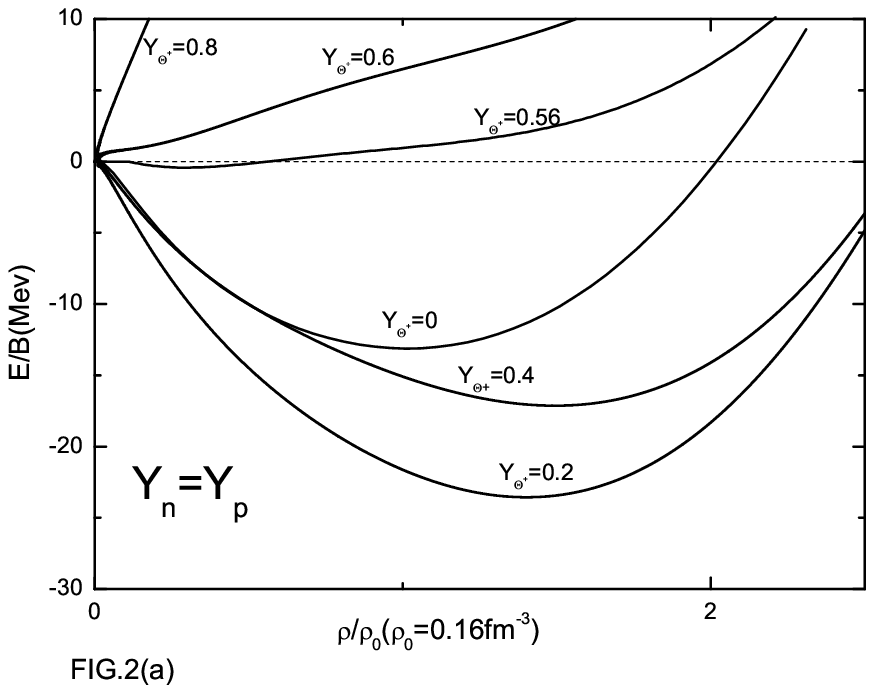}}
\end{center}
\begin{center}
{\includegraphics*[width=0.6\textwidth]{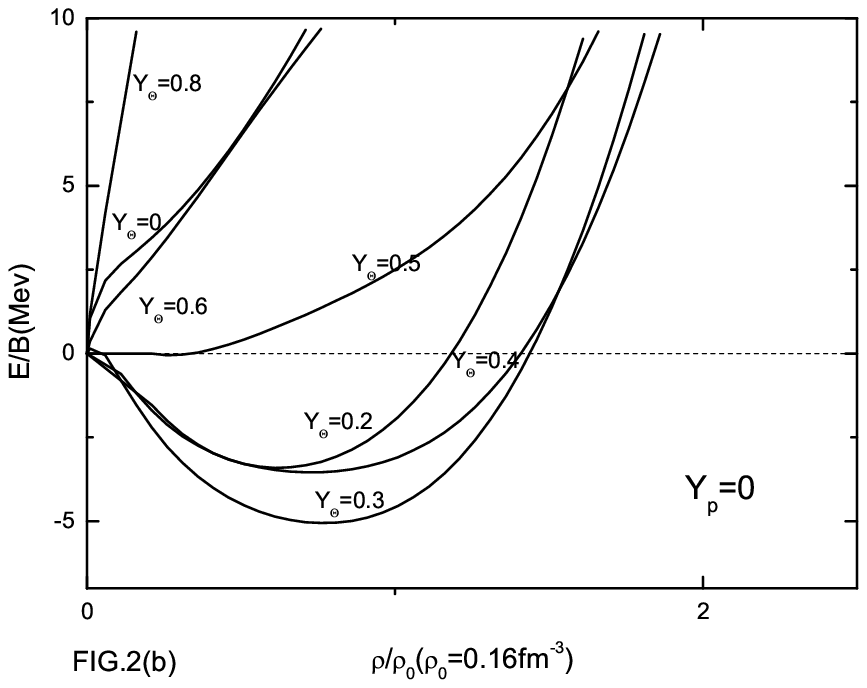}}
\end{center}
\caption{binding energy.}
\end{figure}

\begin{figure}
\begin{center}
{\includegraphics*[width=0.6\textwidth]{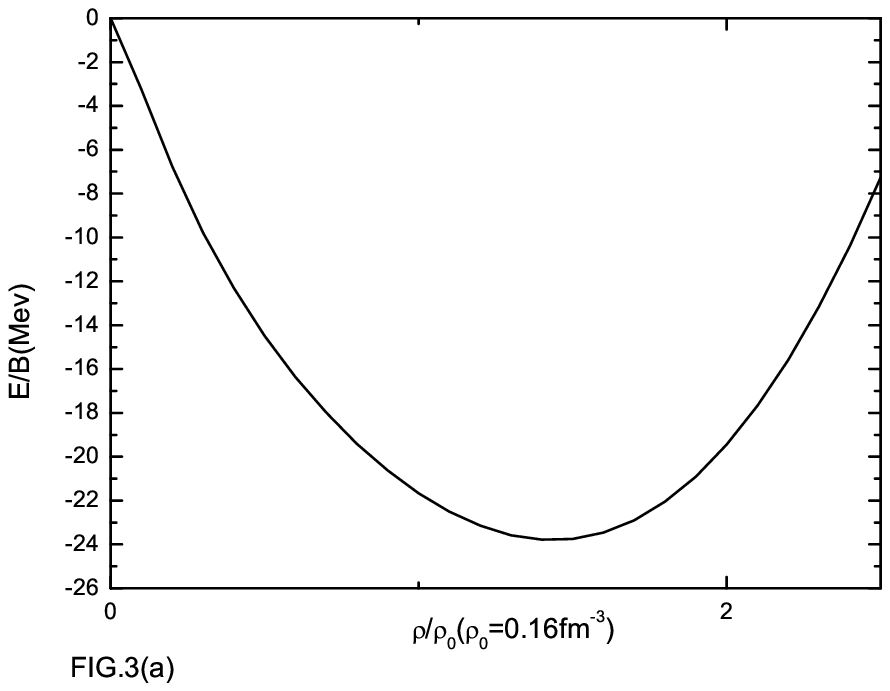}}
\end{center}
\begin{center}
{\includegraphics*[width=0.6\textwidth]{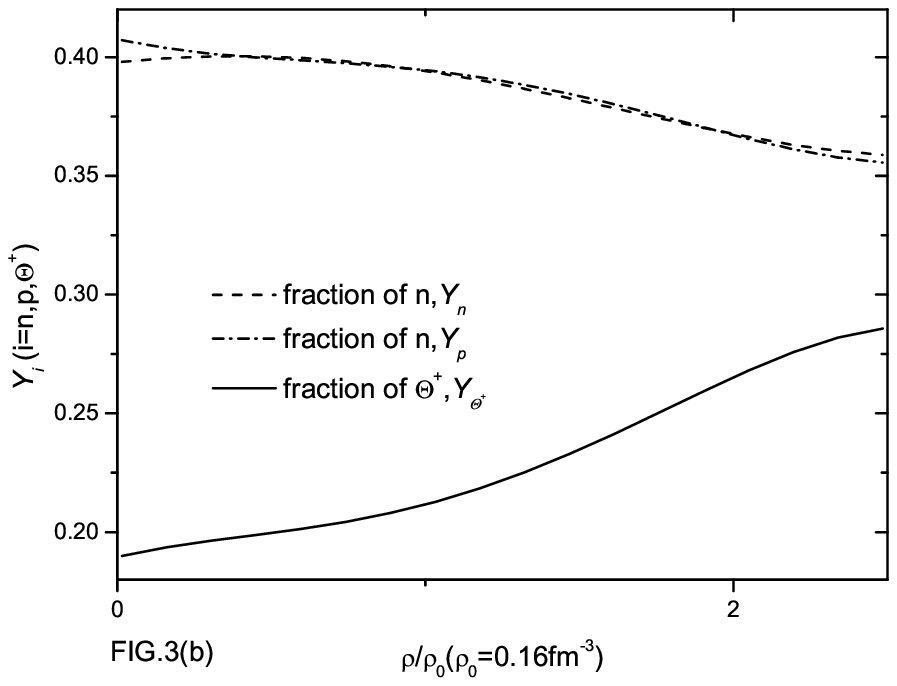}}
\end{center}
\caption{FIG.3(a) shows the minimum of the binding energy.FIG.3(b)
shows the fractions of neutrons,protons and pentaquark $\Theta^+$}
\end{figure}

\end{document}